\documentclass[12pt]{article}
\usepackage{graphicx}
\usepackage{comment}
\newcommand{\age}{\;\raisebox{-.3ex}{$\stackrel{>}{\scriptstyle \sim}$}\;}

\newcommand{\pfrac}[2]{\left(\frac{#1}{#2}\right)}

\newcommand{\aj} {Astron.~J.}
\newcommand{\aap}{Astron.~Astrophys.}
\newcommand{\apj}{Astrophys.~J.}
\newcommand{\apjl}{Astrophys.~J.~Lett.}

\newcommand{\nat}{Nature}
\newcommand{\icarus}{Icarus}

\begin{document}

\centerline{\Large \bf Growing the Gas Giant Planets by}
\centerline{\Large \bf the Gradual Accumulation of Pebbles}

\bigskip
\bigskip

\centerline{\large \bf Harold F$.$ Levison$^{1}$, Katherine A$.$  Kretke$^{1}$, and Martin J$.$ Duncan$^{2}$}

\bigskip

\noindent $^{1}$ {\em Southwest Research Institute and NASA Solar
  System Exploration Research Virtual Institute, 1050 Walnut St, Suite
  300, Boulder, Colorado 80302, USA}

\noindent $^{2}$ {\em Department of Physics, Engineering, and Astronomy, Queen's University, Kingston, Ontario K7L 3N6, Canada}

\vspace{0.5mm}

\bigskip
\bigskip
\centerline{Prepared for {\it Nature}}
\centerline {\today}

\clearpage

\noindent {\bf It is widely held that the first step in forming the gas giant
planets, such as Jupiter and Saturn, is to form solid `cores' of roughly 10
M$_\oplus$ \cite{Mizuno.etal.1978,Pollack.etal.1996}.  Getting the cores to
form before the solar nebula dissipates ($\sim\!1-10\,$Myr
\cite{Haisch.etal.2001}) has been a major challenge for planet formation models
\cite{Goldreich.etal.2004,Levison.etal.2010a}.  Recently models have emerged in
which `pebbles' (centimeter- to meter-size objects) are first concentrated by
aerodynamic drag and then gravitationally collapse to form 100 --- 1000 km
objects \cite{Cuzzi.etal.2001, Youdin.Goodman.2005,
Johansen.etal.2007,Youdin.2011a}.  These `planetesimals' can then efficiently
accrete leftover pebbles \cite{Ormel.Klahr.2010} and directly form the cores of
giant planets
\cite{Lambrechts.Johansen.2012,Lambrechts.Johansen.2014}.  This model
known as `pebble accretion', theoretically, can produce 10 M$_\oplus$ cores in
only a few thousand years \cite{Lambrechts.Johansen.2012,Kretke.Levison.2014}.
Unfortunately, full simulations of this process \cite{Kretke.Levison.2014} show
that, rather than creating a few 10 M$_\oplus$ cores, it produces a population
of hundreds of Earth-mass objects that are inconsistent with the structure of
the Solar System.  Here we report that this difficulty can be overcome if
pebbles form slowly enough to allow the planetesimals to gravitationally
interact with one another. In this situation the largest planetesimals have
time to scatter their smaller siblings out of the disk of pebbles, thereby
stifling their growth.  Our models show that, for a large, and physically
reasonable region of parameter space, this typically leads to the formation of
one to four gas giants between 5 and 15 AU in agreement with the observed
structure of the Solar System.}

Our models consist of a series of computer simulations that follow the
evolution of a population of objects in a disk around the Sun.  The solar
composition disk has surface density distribution $\Sigma \propto r_{\rm
AU}^{-1}$, where $r_{\rm AU}$ is the distance from the Sun, and consists of
both gas and solids.  We assume that an initial population of planetesimals,
which follow the surface density of the disk, form quickly and thus exist at
the beginning of our simulations.  These planetesimals contain only a small
fraction of the mass of solids available (a free parameter of our models).
Pebbles either also existed at the beginning of the simulation, or are allowed
to form over some period of period of time (again a free parameter) starting at
the beginning of the calculation.  The evolution of this system is followed
numerically and includes the effects of gravitational interactions,
interactions between bodies in the disk and the gas (although
nebular tidal migration is neglected), accretion (including enhancements
due to the aerodynamic drag on pebbles; e.g$.$
Ref$.$~\cite{Lambrechts.Johansen.2012}), and collisional fragmentation.
Details of our setup and numerical techniques are described in the {\it Methods
Section}.

\begin{figure}
	\includegraphics{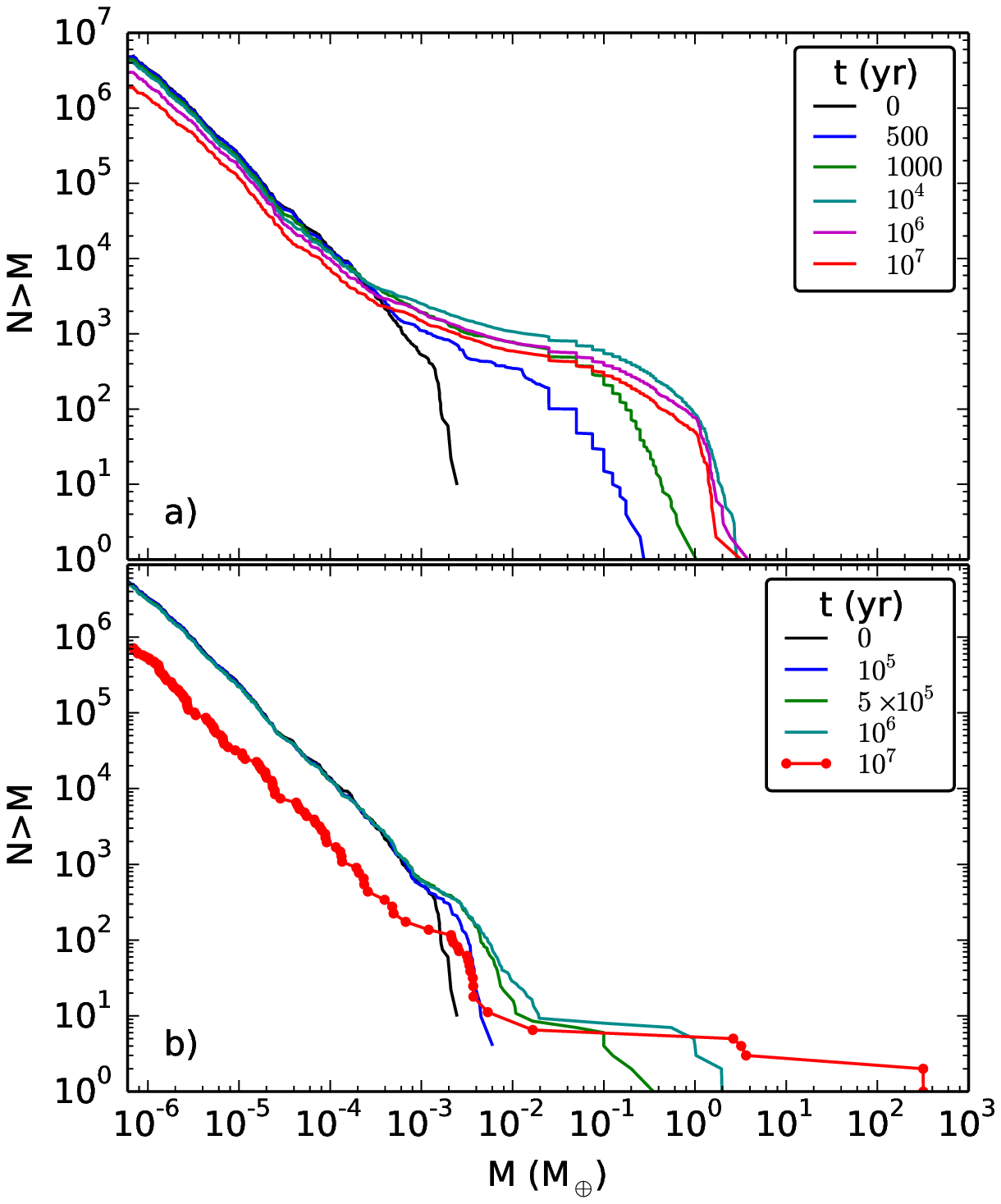}
\caption{ {\bf The cumulative mass distribution
of planetesimals and embryos.} The growth of planetary embryos in our
simulation as illustrated by cumulative mass distributions shown at various
times (indicated by color; see the legend).  {\bf a)} An example of standard
pebble accretion --- all pebbles were in existence at the beginning of the
simulation.  This is similar to Ref.~\cite{Kretke.Levison.2014}. {\bf b)} An example
with the same disk parameters as (a), but where the pebbles formed over the
lifetime of the disk.  This system formed two gas giants and three icy planets
with masses between 2 and $4\,M_{\oplus}$.  The icy planets are on crossing
orbits at the end of the simulation (10 Myr) and thus their number and masses
would likely change as the calculation were run to completion.  \emph{A movie
version of this figure is available in the supplementary material.}
\label{fig:NofM}
}
\end{figure}

In Figure~\ref{fig:NofM} we present the results of two different
simulations that employ the same parameters except for the method of
pebble formation.  In the first (Figure~\ref{fig:NofM}a), we assume
that the pebbles are leftovers of planetesimal formation and thus they
exist at the beginning of the simulation.  Note the fast accretion
times --- embryos (defined to be objects that grow
  significantly) evolved from roughly the mass of Pluto to that of
the Earth in only 1000 years!  However, even though Earth-mass embryos
form quickly, this simulation does not reproduce the Solar System
because rather than forming a few cores, it creates $\sim\!100$
Earth-mass objects.  These objects subsequently scatter one another to
high-eccentricity, high-inclination orbits, stalling growth. This
result is clearly inconsistent with observations.
Ref$.$~\cite{Kretke.Levison.2014} finds that this is a generic outcome
for this type of pebble accretion simulations and thus it could not
have occurred in the Solar System.  It is indicative of the
  results one would achieve if the dynamical interactions between
  planetesimals is neglected (which we refer to as the standard
  model).

In Figure~\ref{fig:NofM}b, we employ a substantial modification to the pebble
accretion theory where we couple an extended timescale for pebbles
formation (a conjecture supported both by the observed presence of cm to mm-sized
grains, which are expected to drift rapidly out of protoplanetary disks, in disks of a wide variety of ages; e.g$.$
Ref.\cite{Ricci.etal.2010} and theoretical models of pebble formation
\cite{Birnstiel.etal.2012}) to the dynamical stirring due to planetesimal
self-gravity (so-called {\it viscous stirring} \cite{Stewart.Wetherill.1988}).
The behavior of this simulation is radically different from that in
Figure~\ref{fig:NofM}a in that, rather than producing a large population of
Earth-mass objects, a few giant planets form.  Only 5 objects grow to
$1\,M_{\oplus}$ or larger in this simulation, and there are two gas giants at
$10\,$Myr.  Also, note that the timescale for growth is very different in the
two calculations.  The first Earth-mass object grows in 1000 years in the
standard pebble accretion run, as opposed to taking slightly over 400,000 years
in this simulation.  In the latter case, the growth time is determined by the
rate at which pebbles were created.

The outcomes of the two simulations are remarkably different because,
in the Figure~\ref{fig:NofM}b simulation only the most massive embryos
are able to accrete a significant amount of pebbles.  This is due to
the fact that the embryos grow slowly enough in this model that they
can gravitationally interact with one another as they accrete the
pebbles.  This viscous stirring has two effects on the
embryo's accretion rates.  First, it increases the relative velocities
and thus decreases the capture cross-section (\emph{c.f. Methods
  Section}, Eq$.$~\ref{eq:R_C}).  More importantly, encounters between
embryos lead to increases in the inclinations, $i$, of the embryos and
thus the distances they travel above and below the disk
midplane. Once the inclinations of the embryos become larger
  than that of the aerodynamically damped pebbles, the embryos spend
much of their time above or below the location where the bulk of the
pebbles lie.  This effectively starves them and stifles their
  growth.  Due to the role of viscous stirring in determining the
outcome in these simulations, we refer to this process as {\it
viscously stirred pebble accretion (VSPA)}.

\begin{figure}
	\includegraphics{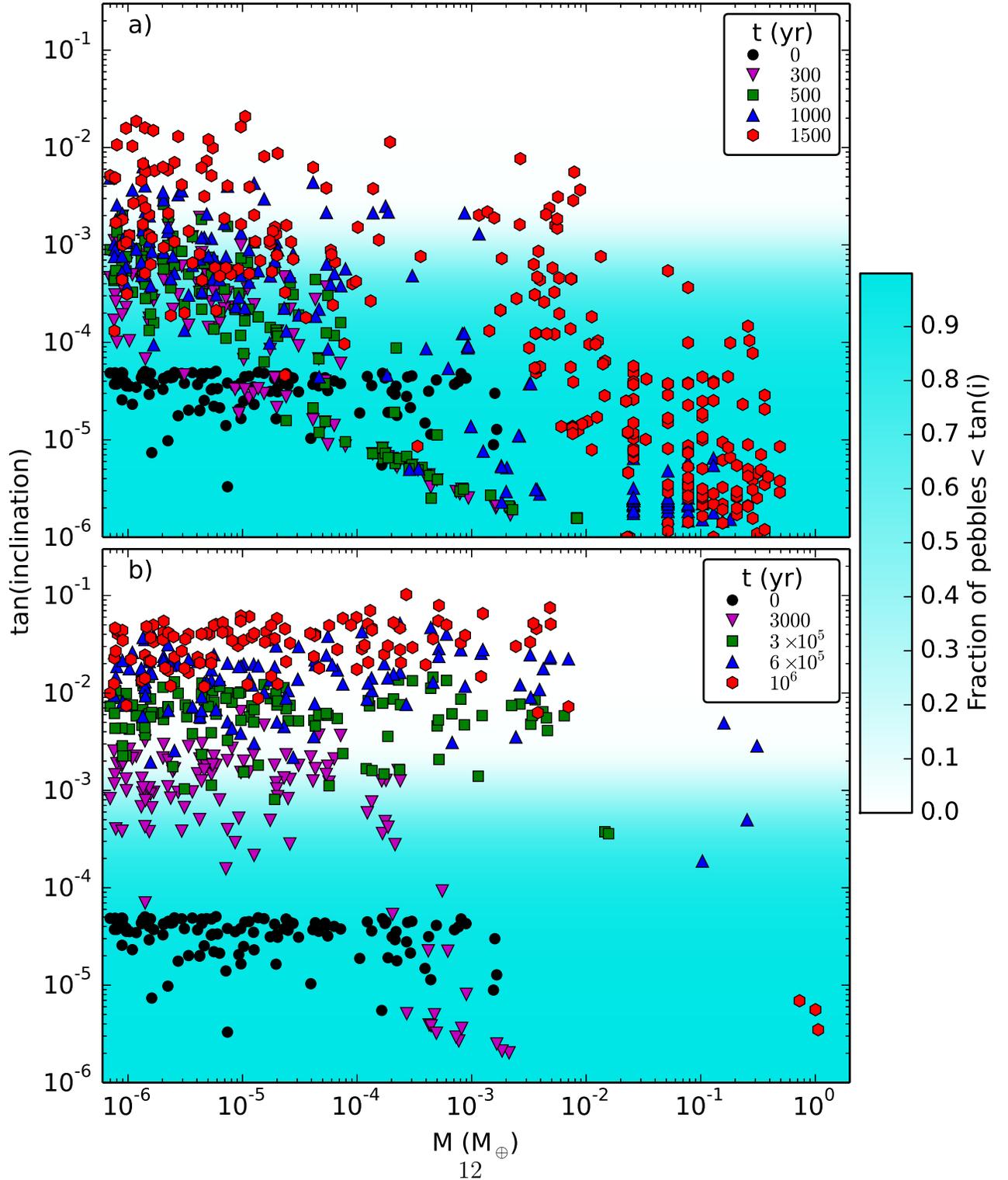}
\caption{ {\bf The vertical distribution of pebbles and
embryos.} A comparison between the vertical distribution, as represented by
$\tan{(i)}$, of pebbles and embryos in the simulations shown in
Figure~\ref{fig:NofM}.  The embryos are shown as dots in the figure, where
their color indicates the time within the simulation (see legend).  Since
objects grow at different times within the disk, we only show embryos from a
narrow annulus ($11.5$ -- $13.5\,$AU).  The time averaged cumulative
inclination distribution of the pebbles is shown by the cyan gradient. {\bf a)}
The run shown in Figure~\ref{fig:NofM}a, {\bf b)} Our fiducial VSPA simulation
from Figure~\ref{fig:NofM}b. \emph{A movie version of this figure is available
in the supplementary material.}
\label{fig:Mz}
}
\end{figure}

Figure~\ref{fig:Mz} illustrates this effect.  In the standard pebble
accretion model, the system inclinations remain small and thus all the
planetesimals can grow.  This because the timescale for viscous
stirring is longer than that of the growth.  In this case it can be
shown (\emph{c.f. Methods Section}) that $dM_e/dt \propto M_e^\beta$
and $\beta \ll 1$.  The small value of $\beta$ implies that smaller
embryos can catch up with larger ones, leading to a population of
like-sized objects.  

However, the dynamical evolution of the system is very different in
the VSPA simulation.  Here viscous stirring can act thereby leading to
an increase of inclinations.  The magnitude of this increase is set by
the mass of the largest embryo --- note the increase with time of both
the inclination of the smaller embryos and the largest embryo mass in
Figure~\ref{fig:Mz}b.  Initially the inclinations of the embryos are
smaller than those of the pebbles, but by 3000 years, most of the
smaller embryos begin to spend a significant amount of time
above or below the pebbles.

The time at which embryos are excited out of the swarm of pebbles can
be estimated with a simple calculation.  The inclinations of the
pebbles are set by the balance between turbulent excitation and
aerodynamic drag in the disk.  For our example, the average pebble
inclination is 0.0016 radians.  Given enough time, a population of
embryos will stir one another to the point where their relative
velocities are comparable to their surface escape velocities.  Since $i \sim
v_{rel}/v_c$, even if we assume objects do not grow, our population of
planetesimals should reach inclinations of $\sim\!0.06$ ---
significantly larger than that of the pebbles.  As a result, this
effect should be important during much of our simulation.  Indeed,
Figure~\ref{fig:Mz} shows that most of the embryos in the system are
in this state after only $\sim\!3000$ years.

The exception is the most massive embryos
(Figure~\ref{fig:Mz}b), which tend to have low inclinations
as a result of a combination of gravitational interactions with
smaller planetesimals (so-called dynamical friction
\cite{Stewart.Wetherill.1988,Ida.1990}) and with the gas (so-called
Type~I inclination damping \cite{Ward.1997}).  As a result, the larger
embryos grow relatively quickly while the smaller embryos grow very
slowly, if at all. Recall that the standard model of pebble accretion
formed a large number of planets because $\beta < 1$, which allowed
smaller embryos to catch up with larger ones.  Figure~\ref{fig:Mdot}
shows the relationship between $dM_e/dt$ and $M_e$ in our fiducial
VSPA run as the system evolves. At early times, $\beta$ is slightly
larger than 1, leading to a few embryos becoming dominate in the
population. When the largest objects in the system have masses between
$\sim\!8 \times 10^{-3}\,M_\oplus$ and $\sim\!0.02\,M_\oplus$ their
equilibrium inclinations decrease, leading to a spurt of growth where
$\beta$ is large ($\sim 4$).  This allows a small number of embryos to
become separated in mass from the rest, and explains why only a small
number of embryos become massive enough to become giant planet cores.
Indeed, for the small embryos, $dM_e/dt$ decreases with time because
of their increasing inclinations (see Figure~\ref{fig:Mz}).  However
for $M_e \age 0.02\,M_\oplus$, $\beta$ becomes small again as the
strongly damped, large embryos accrete in the same manner as in the
standard pebble accretion scenario.  This last phase is important
because it forces these proto-cores to have similar masses as they
grow.  Thus, they all reach the mass where they can directly accrete
gas at roughly the same time.  This last phase might also explain why
the four cores of the giant planets likely originally had similar
masses.  Two gas giants grow in our fiducial simulation (Jupiter and
Saturn?).  By varying parameters in the models, our systems produced
between 0 and $4$ gas giants.  Results of the simulations can be found in Extended Data Tables 1 and 2.

There is another constraint that any model of giant planet formation
in the Solar System must satisfy in order to be considered a success.
The distribution of small bodies in the outer Solar System indicates
that the orbits of the giant planets moved substantially after they
formed \cite{Fernandez.Ip.1984,Malhotra.1995,Tsiganis.etal.2005}.  In particular, Uranus
and Neptune likely formed within 15 or $20\,$AU of the Sun and were
delivered to their current orbits by either a smooth migration
\cite{Malhotra.1995} or a mild gravitational instability
\cite{Tsiganis.etal.2005}.  Both processes require that a population
of planetesimals existed on low-eccentricity orbits beyond the giant
planets after the planets finished forming.  This population must
have: a) not formed planets, and b) survived the planet formation
process relatively unscathed.  To evaluate this constraint in our
simulations, we placed 5 planetesimals with radii similar to Pluto
($s=1350\,$km) on circular, co-planar orbits with semi-major axes
between 20 and $30\,$AU.  None of these objects grew in our fiducial
simulation and they all survived on orbits with eccentricities less
than 0.07 (see the {\it Methods Section} for how this varies
with disk parameters).  Therefore, the process of viscously stirred
pebble accretion reproduces the observed structure of the outer Solar
System: two gas giants, a few icy planets, and a disk of planetesimals
into which the ice giants can migrate.

\section*{Acknowledgements}
This work was supported by an NSF Astronomy and Astrophysics Research Grant (PI Levison).  We would like to thank A.~Johansen, M.~Lambrechts, A.~Morbidelli, D.~Nesvorny, and C.~Ormel for useful discussions.

\section{Author Contributions}
H.F.L.~and K.A.K.~jointly conceived of the paper and carried out the bulk of the numerical and semi-analytic calculations.  M.D.~developed a semi-analytic model of viscous stirring and growth rates in a population distribution.
All authors contributed to the discussion of the results and to the crafting of the manuscript.

\section*{Author Information}
The authors declare no competing financial interests.
Correspondence and requests for materials should be addressed to hal@boulder.swri.edu

\section*{Figure Legends}

\begin{figure}
	\includegraphics{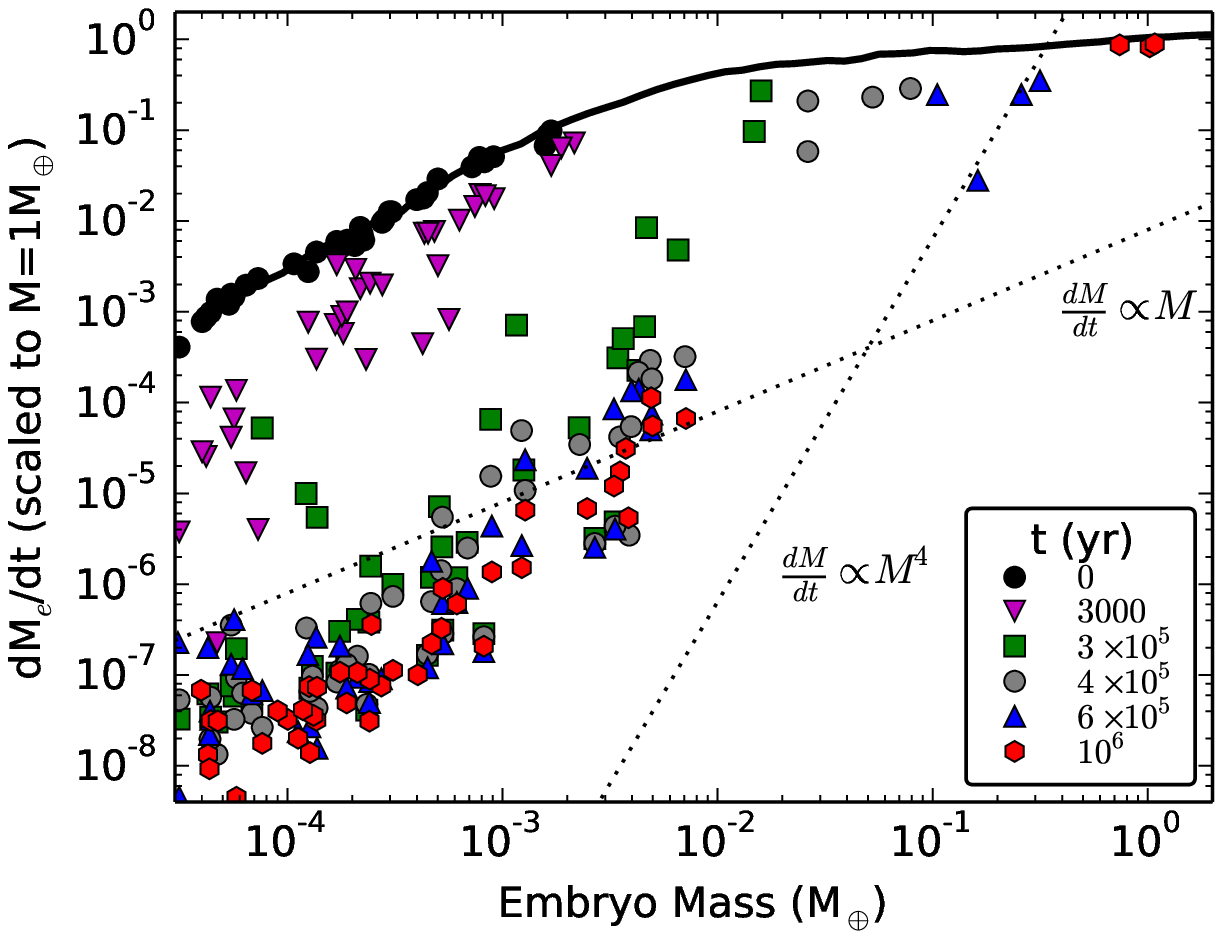}
\caption{  {\bf Embryo growth rate as a function of
mass.} The temporal evolution of the relationship between embryo growth rate
($dM_e/dt$) and mass ($M_e$).  The dots show our fiducial VSPA simulation
(Figure~\ref{fig:NofM}b), where their color indicates the time within the
simulation (see legend). The solid curve is from the standard pebble accretion
run in Figure~\ref{fig:NofM}a.  These values were calculated using procedures
described in the {\it Methods Section}.  The behavior of the system is
determined by the slope of the data in the figure, $\beta$ (note that this is a
log-log plot) at each time.  For reference, the dotted lines show $\beta = 1$
and $\beta = 4$.  For $\beta < 1$, small embryos can catch up with larger ones
leading to a population of like-sized objects.  For $\beta > 1$ the largest
embryos run away from their smaller siblings.  See \emph{Methods} section for more details.
\label{fig:Mdot}
}
\end{figure}

\section*{Methods}
\subsection*{Computational Methods: LIPAD}

For this work we have employed a particle based Lagrangian code (known
as LIPAD, {\bf L}agrangian {\bf I}ntegrator for {\bf P}lanetary {\bf
  A}ccretion and {\bf D}ynamics) that can follow the
dynamical/collisional/accretional evolution of a large number of
planetesimals through the entire growth process to become planets.
For full details about the code and extensive test suites see
Ref.~\cite{Levison.etal.2012}. For details about how pebble accretion
is implemented in the code see Ref.~\cite{Kretke.Levison.2014}, but we
summarize the most relevant attributes of the code here.

LIPAD is built on top of of the N-body integrator SyMBA
\cite{Duncan.etal.1998}.  In order to handle the very large number of
sub-kilometer objects required by many simulations, LIPAD utilizes a
concept known as {\it tracer particles}. Each tracer represents a
large number of small bodies with roughly the same orbit and size, and
is characterized by three numbers: the physical radius, the bulk
density, and the total mass of the disk particles represented by the
tracer.  LIPAD employs statistical algorithms that follow the
dynamical and collisional interactions between the tracers. When a
tracer is determined to have been struck by another tracer, it is
assigned a new radius according to the probabilistic outcome of the
collision based on a fragmentation law by
Ref.~\cite{Benz.Asphaug.1999}, using the ice parameters $Q_0 = 7\times
10^7\,{\rm erg\,g}^{-1}$, $B=2.1\,\rm{erg\,cm}^3\,\rm{g}^{-2}$,
$a=-0.45$ and $b=1.19$.  This way, the conglomeration of tracers
represents the size distribution of the evolving planetesimal
population.  In this work, we do not allow our pebbles to collisionally grow
or fragment, therefore particles below 1~km in size are not involved
in the collisional cascade.  

LIPAD also includes statistical algorithms for viscous stirring, dynamical
friction, and collisional damping among the tracers. The tracers mainly
dynamically interact with the larger planetary mass objects via the normal
$N$-body routines, which naturally follow changes in the trajectory of tracers
due to the gravitational effects of the planets and {\it vice versa}.  LIPAD is
therefore unique in its ability to accurately handle the mixing and
redistribution of material due to gravitational encounters, including gap
opening, and resonant trapping, while also following the fragmentation and
growth of bodies.  Thus, it is well suited to follow the evolution of a
population of embryos, planetesimals, and pebbles while they gravitationally
interact to form planets.

The pebble accretion model follows the prescription in
Ref.~\cite{Ormel.Klahr.2010} and is described in detail in
Ref.~\cite{Kretke.Levison.2014}.  An overview of the physics follows
in the next section.

In the simulations presented here we do not allow the growing planets
to migrate via type-I migration \cite{Ward.1997}, although we do
include type-I eccentricity damping \cite{Papaloizou.Larwood.2000}.
We also neglect type-II migration.  We calculate the aerodynamic drag
on all bodies using the formalism of Ref.~\cite{Adachi.etal.1976}.

Additionally, as we are interested in the gross evolution of a system after the
formation of a potential giant planet core, we have added a simple optional
prescription allowing cores to accrete gas envelopes.  In order to accrete gas
the core size must be above a critical value, which depends on the mass
accretion rate of solids onto the core.  We follow Ref.~\cite{Rafikov.2006} to
determine when core masses are above the critical value given their current
mass accretion rate (assuming a grey opacity of 0.02 times that of the
ISM \cite{Pollack.etal.1996}).  If this criteria is met then we grow the
planet of mass $M_p$ on the Kelvin Helmholtz timescale ($t_{\rm KH}$), so that
\begin{equation}
\dot M_g = \frac{M_p}{t_{\rm KH}}. \nonumber
\end{equation}
Following Ref.~\cite{Ida.Lin.2008a} we approximate this timescale as 
\begin{equation}
t_{\rm KH} = 10^9 \pfrac{M_p}{M_\oplus}^{-3} {\rm yrs}. \nonumber
\end{equation}
We limit gas accretion to the Bondi accretion rate,
\begin{equation}
\dot M_{\rm g, max} = \frac{4\pi\rho_g G^2 M_p^2}{c_s^3}, \nonumber
\end{equation}
where $\rho_g$ is the gas density and $c_s$ is the local sound speed.  
We note that a major uncertainty in these models is the envelope opacity, which
is dominated by the poorly constrained properties of the dust.  Varying the
opacity dramatically alters both the size of the critical core mass, and
perhaps more strikingly, the Kelvin-Helmholtz time.  Additionally, we
arbitrarily cut off gas accretion when a planet reaches one Jupiter mass
instead of  including any physics to reduce the accretion of gas after gap
opening (e.g.~Ref.~\cite{DobbsDixon.etal.2007}). Furthermore, we do not include
the fact that pebbles will cease accreting onto planetesimal cores once the
cores begin perturbing the disk \cite{Lambrechts.etal.2014}.  This effect may
prove important in causing some cores to accrete gas sooner, or in allowing
cores to grow to larger sizes without accreting gas (creating true Uranus and
Neptune analogs).  Therefore the end masses of giant planets should be viewed
with these caveats in mind.

We also note that while this model includes fragmentation, we have found that
the production, and subsequent sweep-up of pebbles (as considered by
Ref.~\cite{Chambers.2014}) is relatively unimportant in these scenarios.  This
is due to the assumed relatively low initial mass of planetesimals in our
simulations (17 $M_\oplus$ between 4 and 15 AU in our fiducial simulation).

\subsection*{Pebble Accretion} 

Here we present an argument, based on
Refs.~\cite{Ormel.Klahr.2010,Lambrechts.Johansen.2012}, about why
pebbles are effectively accreted by growing planets.  If the stopping
time ($t_s$) of a pebble is comparable to the time for it to encounter
a growing embryo, then it can be deflected out of the gas stream and
lose enough orbital energy to become gravitationally bound to the
embryo. After capture, the pebble spirals inward due to aerodynamic
drag and is accreted.  In this case the collisional cross section for
accretion is

\begin{equation}
	\sigma_{\rm peb} \equiv \pi \pfrac{4 G M_e t_s}{v_{rel}}\exp\left[-2\pfrac{t_sv_{rel}^3}{4G M_e}^{\gamma}\right],
	\label{eq:R_C}
\end{equation}
where $M_e$ is the mass of the embryo, $v_{rel}$ is the relative
velocity between the pebble and embryo, $G$ is the gravitational
constant, and $\gamma = 0.65$ \cite{Ormel.Klahr.2010}.  When the
Stokes number of a pebble ($\tau\equiv t_s \Omega_K$, where $\Omega_K$
is the local Keplerian frequency) is near unity, this capture radius
can be $10^7$ times larger than the physical cross section alone ($\pi
R_e^2$, where $R_e$ is the radius of the embryo) for Earth-sized
planets in the region inhabited by the giant planets.

This process can cause an isolated $1000\,$km object to grow into a 10
M$_\oplus$ core in only a few thousand years
\cite{Lambrechts.Johansen.2012,Kretke.Levison.2014}.  To zeroth order, the
accretion rate of an embryo growing by pebble accretion is $dM_e/dt = f_{\rm
ac} \mathcal{R}$, where $f_{\rm ac}$ is the fraction of pebbles that drift into
an embryos orbit that will be accreted (also called the filtering factor, which
is a function of $\sigma_{\rm peb}$ and the spatial distribution of pebbles),
${\mathcal R} = 2\pi r \Sigma_{\rm peb} v_{\rm rad}$ is the rate at which
pebbles are fed to the embryo, $\Sigma_{\rm peb}$ is the surface density of
pebbles at heliocentric distance $r$, and $v_{\rm rad}$ is the radial velocity
of the pebbles due to aerodynamic drag.  Pebbles drift at a velocity
\begin{equation} 
	v_{\rm rad} =  -2 \frac{\tau}{\tau^2+1} \eta v_c, \nonumber
\end{equation} 
where $\eta$ is a dimensionless parameter related to the
pressure gradient of the gas disk \cite{Nakagawa.etal.1986} and is on the order
of a few times $10^{-3}$, and $v_c$ is the local circular velocity of the
embryo.  For $\tau \sim 1$, a pebble can spiral through the disk is only a few
hundred years.  These large values of $v_{\rm rad}$ can lead to huge accretion
rates and giant planet cores can potentially grow quickly.

\subsection*{Pebble Formation Model}

In this work we utilize a simple prescription to convert dust into pebbles over
time.  We assume that the gas disk with mass $M_g$ exponentially decays over a
timescale $t_g$, so that
\begin{equation}
	\dot{M}_g = -\frac{M_g}{t_g}.
\end{equation}
Pebbles (with total mass $M_{\rm peb}$) are formed at a rate proportional to the square of the dust mass ($M_d$) such that
\begin{equation}
	\dot{M}_{\rm peb} = k M_d^2, 
\end{equation}
where $k$ determines the rate of pebble formation.  The dust in the
disk will be lost as the gas disk evolves and as pebbles form,
  yielding a dust evolution of
\begin{equation}
	\dot{M}_d = \dot{M}_g \frac{M_d}{M_g}-\dot{M}_{\rm peb}.
\end{equation}
This leads to a total rate of pebble production of
\begin{equation}
	\dot{M}_{\rm peb} = \frac{\kappa}{t_g} M_{d,0} \frac{1}{(1+\kappa)^2}\left(\frac{1}{\exp(t/t_g)-\kappa/(1+\kappa)}\right)^2,
\end{equation}
where $\kappa \equiv k t_g M_{d,0}$.  Motivated by observations of
disks, we take the time constant for pebble production to be large,
with a median production timescale near $1\,{\rm Myr}$.  We assume
that all of the pebbles are produced in 3 Myr.  For simplicity, we
assume that pebbles are randomly created throughout the disk according
to the surface density, and the size of the pebbles is determined by
the assumed $\tau$, which is constant in each simulation, but
varies between calculations.

To test the robustness of our result to the specific assumptions of
our pebble formation model, we modified it in two different ways for
the runs shown in the Extended Data Tables.  In some
runs we allow the pebbles to grow as they drift inwards by sweeping up
dust.  Their growth rate is thus
\begin{equation}
	\frac{dm}{dt} = \rho_d \pi r_{\rm peb}^2 v_{\rm rel},
\end{equation}
where $\rho_d$ is the dust mass (which is depleted as pebbles are formed, as
dust is swept up, and as the gaseous disk evolves), $r_{\rm peb}$ is the size
of the pebble, and $v_{\rm rel}$ is the relative velocity between the pebble
and the dust (which is assumed to be perfectly coupled to the gas).

In other runs, we assume a model of pebble formation consistent with the inside
out manner suggested by Ref.~\cite{Birnstiel.etal.2012}.  We used the
implementation as described in Ref.~\cite{Lambrechts.Johansen.2014} and had a
wave of pebble formation which moves outwards as a function of time.  In
particular, the radius at which pebbles are generated ($r_g$) at time $t$ is 
\begin{equation}
r_g(t) = \left(\frac{3}{16}\right)^{1/3} (G M_*)^{1/3} (\epsilon_d Z_0)^{2/3} t^{2/3},
\end{equation}
where $Z_0$ is the metallicity of the disk and $\epsilon_d$
encapsulates the efficiency of particle growth.  We modify
$\epsilon_d$ to adjust the timescale of pebble formation.  We found
that so long as the coagulation coefficient is small enough that the
timescale for pebble formation is long, we can get results that are
generally similar to our fiducial model in which pebbles form
randomly.

\subsection*{Model Parameters and Simulation Statistics}

Since we are interested in building the gas giant planets, we study the
growth of planetesimals spread from $4$ to $15\,$AU.  As we integrate the
simulation forwards, pebbles are formed between 4 and $30\,$AU.  Each system
was evolved $10\,$Myr using LIPAD.  Extended Data Tables 1 and 2 list the 42
simulations that we have completed in our investigation of giant planet
formation.  The eight important parameters that were varied are:

\begin{enumerate}

\item{} The surface density of the gas disk at $1\,$AU, $\Sigma_0$.  In all our
simulations we used a surface density distribution of $\Sigma = \Sigma_0 r_{\rm
AU}^{-1}$\cite{Andrews.etal.2010}, where $r_{\rm AU}$ is the heliocentric distance in AU.  For the
fiducial simulation, $\Sigma_0 = 7200$ g/cm$^2$, which is 4 times the surface
density of the minimum mass solar nebula at that location \cite{Hayashi.1981}.
The gas surface density decreases exponentially
with a timescale of $2\,$Myr \cite{Haisch.etal.2001, Hernandez.etal.2009}.

\item{} We employ a flaring gas disk with a scale height $h =
  0.047~r_{\rm AU}^{q}~{\rm AU}$.  We used two values for $q$: 1.25
  from Ref$.$~\cite{Hayashi.1981} and $9/7$ from
  Ref$.$~\cite{Chiang.Goldreich.1997}.  The later was used in our fiducial run.

\item{} The size of the largest planetesimal, $s_{\rm max}$. We draw
  the initial planetesimals from a distribution of radii, $s$, of the
  form $dN/ds \propto s^{-4.5}$ such that $s$ is between $100\,$km
  \cite{Morbidelli.etal.2009} and $s_{\rm max}$. For our fiducial simulation
  $s_{\rm max} = 1350\,$km, which is slightly larger than Pluto.

\item{} The fraction of solids in the disk that are initially
  converted into planetesimals, $f_{pl}$.  For all simulations, we
  assume a solid-to-gas ratio of 0.01, which includes the contribution
  of water ice \cite{Lodders.2003}, and $f_{pl}$ of the solids are in planetesimals while $(1-f_{pl})$ are in pebbles.  For the fiducial simulation,
  $f_{pl}= 0.1$,  which, if extended to
$30\,$AU, is consistent with the mass of planetesimals needed to subsequently
deliver the giant planets to their current orbits \cite{Tsiganis.etal.2005}.

\item{} The initial Stokes number of a pebble, $\tau$.  Note that as
  pebbles spiral toward the Sun, their Stokes number changes
  (generally decreases) even when it is assumed that they do not
  grow. For our fiducial simulation $\tau = 0.6$.

\item{} The strength of inclination and eccentricity damping due to
  the gravitational interaction with the gas disk, $c_e$.  We employ
  the techniques in Ref$.$~\cite{Levison.etal.2010a}, which are based on
  Ref$.$~\cite{Papaloizou.Larwood.2000} and allow for both radial
  migration and inclination/eccentricity damping.  The strength of
  both processes can be adjusted by varying two dimensionless
  parameters: $c_a$ and $c_e$, respectively.  All our simulations have
  $c_a = 0$, meaning there is no Type~I migration.  For our fiducial
  simulation $c_e = 0.66$.

\item{} LIPAD \cite{Levison.etal.2012}, has the option of allowing
  pebbles to grow by the accretion of dust particles suspended in the
  gas.  The growth rates are calculated using the particle-in-a-box
  approximation assuming the distribution of dust follows that of the
  gas, as described above.  Our fiducial simulation has this option disabled.

\item{} The median time of pebble generation, $t_{\rm gen}$.  For most
  of our simulations, pebble generation follows the evolution of the
  gas disk.  In this case $t_{\rm gen} \sim 0.7\,$Myr.  However, in
  two cases we decreased $t_{\rm gen}$ in order to determine whether
  it affects our results.  We found that the value of $t_{\rm gen}$ is
  not important as long as it is larger than the viscous stirring
  timescale of the embryos.  Times marked with an asterisk indicate
  that pebbles were generated in an inside-out manner as described above.

\end{enumerate}

\noindent Additionally, all the disks are turbulent with $\alpha=4\times 10^{-3}$
\cite{Shakura.Sunyaev.1973}.

The last five columns in the table present the basic
characteristics of our systems.  In particular, $N^{\rm Gas}$ is the
total number of gas giant planets that formed during each run.  It
ranges from 0 to 4.  Occasionally, a giant planet or two was lost from
the system.  Although some of these were ejected due to dynamical
instabilities, the majority were pushed off the inner edge of our
computational domain by their neighbors as the latter accreted gas.
Unfortunately, the inner edge of the domain needs to be relatively
large because it determines the timestep used in the calculation.  The
$N$-body integrator requires that the perihelion passage be temporally
resolved, and since the location of the domain's inner boundary
determines the smallest perihelion, it also sets the timestep.  In
order for the calculations to be practical, we were forced to remove
any object with a perihelion distance smaller than $2.7\,$AU (the
location of the snow line).  The majority of the lost gas giants
  would likely have remained in the asteroid belt or migrated
    into the terrestrial planet region if we had been able to keep
  them in the calculations.  The column labeled $N^{\rm Gas}_f$ gives
the number of gas giants remaining in the system at $10\,$Myr.

Columns (11) and (12) refer to icy objects with masses greater than
$1\,M_\oplus$.  Again, $N^{\rm ice}$ and $N^{\rm ice}_f$ are the total number
of planets that formed and the number remaining at $10\,$Myr, respectively.
Unlike their larger siblings, the majority of the lost icy planets were ejected
from the planetary system, mainly by encounters with growing gas giants.  The
number of lost ice planets illustrates the fact that the systems becomes
violent after the largest cores begin to accrete gas.  The last column
shows whether the 5 planetesimals we placed on circular, co-planar orbits
between 20 and $30\,$AU (as proxies for the planetesimal disk needed for later
planet migration \cite{Tsiganis.etal.2005}) survived at $10\,$Myr.

There are a couple of important caveats that must be taken into account to
properly interpret the results in the Extended Data Tables.  First the
parameter space for our calculations is 8-dimensional, and each calculation
required many weeks of CPU time. This limited the total number that could be
performed.  Thus, it was impossible to cover parameter space in a meaningful
way.  Instead, we surgically approached the problem by choosing a couple of
starting locations and varying individual parameters to test their effects.  We
then moved in a particular direction in parameter space if we believed we would
be more likely to produce systems with some desired characteristic.  As a
result, the distribution of our results seen in the table (for example the
number of gas giants) cannot be interpreted as the true distribution that would
result if parameter space were uniformly covered.

Our second concern is the crude methods we used to model the direct accretion
of gas onto the cores to form gas giants.  As discussed above, our
simple model for calculating for the onset of gas accretion and the final mass
of planets likely is missing important physics. These limitations mean that,
while our calculations show the pebble accretion produces the correct number of
giant planet cores on the correct timescales, the details of the evolution of
our systems after the gas giants form should be viewed with skepticism.

\subsection*{Generating Figure~\ref{fig:Mdot}}

The data in Figure~\ref{fig:Mdot} were generated using the following
procedure.  We first needed to construct a high resolution
distribution of pebbles.  This was accomplished by summing the output
steps of our integration over all time. Then at each time plotted, we
noted the mass and orbit of each embryo.  We then generated 1000
clones of each embryo --- each with the same semi-major axis,
eccentricity, and inclination, but with different orbital angles.  For
each clone, we calculated its velocity with respect to the pebbles and
then $\sigma_{\rm peb}$ from Eq.~(\ref{eq:R_C}).  From this we
estimate the instantaneous accretion rate using the particle-in-a-box
approximation, $\dot M_e=\rho\,\sigma_{\rm peb}v_{\rm rel}$ where
$\rho$ is the local mass density of pebbles.  The dots in the figure
show the average of the 1000 clones of the instantaneous accretion
rates for each embryo.

\subsection*{Code Availability} The calculations presented in this
paper were performed using LIPAD, a proprietary software product funded by the
Southwest Research Institute and is not publicly available.  It is based upon
the N-Body integrator SyMBA, which is publicly available at
http://www.boulder.swri.edu/swifter/

\section*{Extended Data Legends}
\setlength{\tabcolsep}{4pt}
\begin{table}
\caption{{\bf Our completed simulations.} Please see the \emph{Methods Section} for a full description of the columns in this table.}
\begin{tabular}{|cccccccc|ccccc|}
\hline
    (1)     & (2)& (3)      & (4)    & (5)   &  (6)    &(7)&    (8)      &   (9)  & (10)  &    (11)      &  (12) &  (13) \\     
  $\Sigma_0$&  q &  $s_{\rm max}$  &  $f_{pl}$ & $\tau$ & $c_e$ & pebbles &  $t_{\rm gen}$ &  $N^{\rm Gas}$ & $N^{\rm Gas}_f$ &  $N^{\rm ice}$ &  $N^{\rm ice}_f$ & Disk ok?\\
            &    &             &          &        &       &  grow?  &              &        &              &             &         &         \\
(g/cm$^2$)  &    &(km)          &          &        &       & (Y/N)   &    (Myr)     &        &              &              &        &      (Y/N)     \\
\hline
\hline
3600 & $9/7$ & 1500  & 0.2  & 0.3 & 1    & N & 0.7 & 4 & 2 & 12 &  0 & Y\\
3600 & $9/7$ & 1500  & 0.2  & 0.3 & 1    & N & 0.5 & 4 & 3 & 14 &  0 & Y\\
3600 & $9/7$ & 1500  & 0.2  & 0.3 & 1    & N & 0.2 & 4 & 3 & 15 &  1 & N\\
3600 & $9/7$ & 1500  & 0.2  & 0.4 & 1    & N & 0.7 & 3 & 1 &  9 &  2 & Y\\
3600 & 1.25  & 1500  & 0.2  & 3   & 1    & Y & 0.7 & 3 & 2 &  9 &  0 & Y\\
4050 & $9/7$ & 1500  & 0.18 & 0.4 & 1    & N & 0.7 & 4 & 2 &  6 &  1 & Y\\
4050 & $9/7$ & 1500  & 0.18 & 0.5 & 1    & N & 0.7 & 3 & 3 &  7 &  0 & Y\\
4050 & $9/7$ & 1500  & 0.18 & 0.6 & 1    & N & 0.7 & 4 & 4 &  9 &  0 & Y\\
4500 & $9/7$ & 1500  & 0.16 & 0.3 & 1    & N & 0.7 & 4 & 3 & 13 &  0 & Y\\
4500 & $9/7$ & 1500  & 0.16 & 0.4 & 1    & N & 0.7 & 3 & 2 &  7 &  1 & Y\\
4500 & $9/7$ & 1500  & 0.16 & 0.5 & 1    & N & 0.7 & 3 & 3 & 13 &  0 & N\\
4500 & $9/7$ & 1500  & 0.16 & 0.5 & 0.33 & N & 0.7 & 4 & 2 &  4 &  0 & Y\\
4500 & $9/7$ & 1500  & 0.16 & 0.5 & 0.1  & N & 0.7 & 0 & 0 &  7 &  6 & Y\\
4500 & $9/7$ & 1500  & 0.16 & 0.6 & 1    & N & 0.7 & 2 & 2 & 15 &  3 & N\\
4500 & $9/7$ & 1500  & 0.16 & 0.7 & 1    & N & 0.7 & 0 & 0 &  7 &  7 & Y\\
4500 & $9/7$ & 1500  & 0.16 & 0.9 & 1    & N & 0.7 & 1 & 0 &  1 &  1 & Y\\
5400 & 1.25  & 1500  & 0.15 & 3   & 1    & Y & 0.7 & 0 & 0 &  5 &  4 & Y\\
5400 & 1.25  & 1500  & 0.15 & 3   & 1    & Y & 0.7 & 0 & 0 &  6 &  6 & Y\\
5400 & 1.25  & 1500  & 0.15 & 3   & 0.66 & Y & 0.7 & 0 & 0 &  5 &  5 & Y\\
5400 & 1.25  & 1500  & 0.15 & 3   & 0.33 & Y & 0.7 & 1 & 1 &  7 &  4 & Y\\
7200 & $9/7$ & 1350  & 0.25 & 0.6 & 1    & N & 0.7 & 4 & 4 & 10 &  2 & Y\\
7200 & $9/7$ & 1350  & 0.15 & 0.6 & 1    & N & 0.7 & 4 & 4 &  8 &  0 & N\\
7200 & 1.25  & 1500  & 0.1  & 1   & 1    & Y & 0.7 & 2 & 2 & 11 &  4 & Y\\
7200 & 1.25  & 1500  & 0.1  & 1   & 0.25 & Y & 0.7 & 3 & 3 &  5 &  1 & Y\\
7200 & 1.25  & 1500  & 0.1  & 1   & 0    & Y & 0.7 & 2 & 2 &  1 &  0 & Y\\
7200 & $9/7$ & 1350  & 0.1  & 0.6 & 1    & N & 0.7 & 2 & 2 &  8 &  2 & Y\\
7200 & $9/7$ & 1350  & 0.1  & 0.6 & 0.66 & N & 0.7 & 2 & 2 &  9 &  3 & Y\\
7200 & $9/7$ & 1350  & 0.1  & 0.6 & 0.66 & N & 0.7 & 3 & 3 &  5 &  1 & Y\\
7200 & $9/7$ & 1350  & 0.05 & 0.6 & 1    & N & 0.7 & 2 & 2 &  2 &  1 & Y\\
7200 & $9/7$ & 1350  & 0.025& 0.6 & 1    & N & 0.7 & 3 & 3 & 13 &  0 & N\\
\hline
\end{tabular}
\label{tab:runs}
\end{table}

\setlength{\tabcolsep}{4pt}
\begin{table}
	\caption{ {More Simulations.} This is a continuation of Extended Data Table 1.  Please see the \emph{Methods Section} for a full description of the columns in this table. }
\begin{tabular}{|cccccccc|ccccc|}
\hline
    (1)     & (2)& (3)      & (4)    & (5)   &  (6)    &(7)&    (8)      &   (9)  & (10)  &    (11)      &  (12) &  (13) \\     
  $\Sigma_0$&  q &  $s_{\rm max}$  &  $f_{pl}$ & $\tau$ & $c_e$ & pebbles &  $t_{\rm gen}$ &  $N^{\rm Gas}$ & $N^{\rm Gas}_f$ &  $N^{\rm ice}$ &  $N^{\rm ice}_f$ & Disk ok?\\
            &    &             &          &        &       &  grow?  &              &        &              &             &         &         \\
(g/cm$^2$)  &    &(km)          &          &        &       & (Y/N)   &    (Myr)     &        &              &              &        &      (Y/N)     \\
\hline
\hline
 900 & $9/7$ & 1500  & 0.8 & 0.1 & 0.66 & N & 0.9* & 0 & 0 & 21 & 15 & Y\\ 
 900 & $9/7$ & 1500  & 0.8 & 0.1 & 0.66 & N & 1.3* & 0 & 0 & 17 & 14 & Y\\ 
1800 & $9/7$ & 1500  & 0.4 & 0.1 & 0.66 & N & 0.9* & 5 & 4 &  8 &  0 & N\\ 
1800 & $9/7$ & 1500  & 0.4 & 0.1 & 0.66 & N & 1.3* & 3 & 3 &  6 &  0 & N\\ 
3600 & $9/7$ & 1500  & 0.2 & 0.1 & 0.66 & N & 0.9* & 7 & 5 & 12 &  0 & N\\ 
3600 & $9/7$ & 1500  & 0.2 & 0.1 & 0.66 & N & 1.3* & 6 & 4 &  9 &  0 & N\\ 
3600 & $9/7$ & 1500  & 0.2 & 0.1 & 0.66 & N & 0.4* & 4 & 3 &  9 &  0 & N\\ 
3600 & $9/7$ & 1500  & 0.2 & 0.6 & 0.66 & N & 0.9* & 5 & 3 &  6 &  0 & N\\ 
3600 & $9/7$ & 1500  & 0.2 & 0.6 & 0.66 & N & 0.4* & 2 & 2 & 11 &  4 & Y\\ 
7200 & $9/7$ & 1500  & 0.1 & 0.6 & 0.66 & N & 0.4* & 3 & 1 & 10 &  1 & N\\ 
7200 & $9/7$ & 1500  & 0.1 & 0.6 & 0.66 & N & 0.9* & 4 & 3 &  6 &  0 & N\\ 
7200 & $9/7$ & 1500  & 0.1 & 0.6 & 0.66 & N & 1.3* & 5 & 3 &  5 &  1 & N\\ 
\hline
\end{tabular}
\label{tab:runs2}
\end{table}

\end{document}